\documentclass{article}

\usepackage[final]{neurips_2024}

\usepackage[utf8]{inputenc} 
\usepackage[T1]{fontenc}    
\usepackage{hyperref}       
\usepackage{url}            
\usepackage{booktabs}       
\usepackage{amsfonts}       
\usepackage{nicefrac}       
\usepackage{microtype}      
\usepackage{xcolor}         
\usepackage{graphicx}
\usepackage{amsmath}
\usepackage{natbib}
\title{SpectraFM: Tuning into Stellar Foundation Models}

\author{%
  Nolan Koblischke\(^*\), Jo Bovy \\
  David A. Dunlap Department of Astronomy and Astrophysics\\
  University of Toronto\\
  50 St. George Street, Toronto, Ontario,
M5S 3H4, Canada \\
  \(^*\)\texttt{nolan.koblischke@astro.utoronto.ca} \\
}

\newcommand{\teff}{\(T_{\mathrm{eff}}\)}
\newcommand{\logg}{log \(g\)}

\newcommand{\astronn}{\texttt{AstroNN}}

\usepackage{xcolor}

\begin{document}

\maketitle

\begin{abstract}
Machine learning models in astrophysics are often limited in scope and cannot adapt to data from new instruments or tasks. We introduce SpectraFM, a Transformer-based foundation model architecture that can be pre-trained on stellar spectra from any wavelength range and instrument. SpectraFM excels in generalization by combining flexibility with knowledge transfer from pre-training, allowing it to outperform traditional machine learning methods, especially in scenarios with limited training data. Our model is pre-trained on approximately 90k examples of synthetic spectra to predict the chemical abundances (Fe, Mg, O), temperature, and specific gravity of stars. We then fine-tune the model on real spectra to adapt it to observational data before fine-tuning it further on a restricted 100-star training set in a different wavelength range to predict iron abundance. Despite a small iron-rich training set of real spectra, transfer learning from the synthetic spectra pre-training enables the model to perform well on iron-poor stars. In contrast, a neural network trained from scratch fails at this task. We investigate the Transformer’s attention mechanism and find that the wavelengths receiving attention carry physical information about chemical composition. By leveraging the knowledge from pre-training and its ability to handle non-spectra inputs, SpectraFM reduces the need for large training datasets and enables cross-instrument and cross-domain research. Its adaptability makes it well-suited for tackling emerging challenges in astrophysics, like extracting insights from multi-modal datasets.
\end{abstract}

\section{Introduction}

In many scientific fields, a rapid increase in data availability has led to the wide-spread use of machine learning methods~\citep{smith_astronomia_2023}. The machine learning models used in research are often restricted to a single task, e.g. to predict the iron content of stars from spectra, and only make accurate predictions on data from the instrument that collected its training set. Foundation models, pre-trained on multiple tasks and data from a variety of sources, have unlocked a versatility that was lacking with `classical' machine learning algorithms~\citep{bommasani_opportunities_2022}. Pre-trained foundation models can be released to scientists and used out-of-the-box, or fine-tuned for specific research tasks, in the hopes of leveraging transfer learning to outperform a machine learning algorithm trained from scratch~\citep{mccabe_multiple_2023}. They are particularly useful when applied to 1) datasets too small to train an effective machine learning model~\citep{walmsley_scaling_2024} and 2) cross-instrument and cross-domain datasets that require a synergistic analysis~\citep{parker_astroclip_2024}.

Both situations appear frequently in astronomy. For example, spectroscopic observations from the James Webb Space Telescope (JWST) with accurate stellar property labels are insufficient in number for a training set, O(\(\sim\)dozens), and may not represent the required diversity of stellar properties to train a useful machine learning model. However, if a foundation model were pre-trained on a much larger dataset of stellar spectra, from other telescopes and synthetic sources, and then fine-tuned on the JWST dataset, it could transfer its knowledge to this new task.

Training on synthetic data for application to real data often leads to poor predictions due to the `synthetic gap' - the differences between the simplified synthetic spectra and the complex observed spectra~\citep{fabbro_application_2018, obriain_cycle-starnet_2021}. The synthetic gap stems from physical assumptions and idealized instruments in synthetic models, leading to systematic biases and reduced prediction accuracy when applied to real-world data. We investigate a potential remedy: fine-tuning the model on a small but well-characterized set of real spectra, allowing the foundation model to adjust to these characteristics of real observational data.

The second situation appears when the same source (e.g. a star or galaxy) has been observed by multiple instruments and in different modalities. APOGEE (Apache Point Observatory Galactic Evolution Experiment) and the Gaia space telescope are two large-scale astronomical surveys \citep{majewski_apache_2017,gaia_collaboration_gaia_2023}. Gaia Data Release 3 has released photometric observations, positions, and motions for more than a billion stars. APOGEE Data Release 17 provides high-resolution infrared spectra for over 650,000 stars, capturing crucial atomic lines necessary for determining the abundances of elements in these stars, vital information for stellar and galactic astrophysics. Recently, machine learning models have combined APOGEE spectra observations with Gaia observations to extract more information about stellar properties than what can be done with either alone \citep{cantat-gaudin_uniting_2024,laroche_closing_2024}. 

\cite{leung_towards_2023}, hereafter LB23, developed a proof-of-concept Transformer-based foundation model for stars observed by Gaia and other sources. The single model trained by LB23 can predict stellar properties like temperature, gravity, and chemical composition from low-resolution Gaia spectra or other properties, generate synthetic spectra from stellar parameters, and reconstruct missing spectral regions, which demonstrates a versatility in handling astronomical data.

Furthermore, work is underway to create a single database that encompasses dozens of surveys and modalities with the intention of developing an astronomy-wide foundation model \citep{MultimodalUniverse}. However, the prototype from LB23 cannot be scaled up to accomplish this feat, since it only accepts tabular data and lacks the ability to work with other modalities, like images, time-series measurements, and high-resolution spectra. We adapt the LB23 foundation model to work with stellar spectra from any wavelength, not just the low-resolution Gaia spectra it was trained on. Extracting information from other wavelength ranges and higher resolutions is critically important for stellar astrophysics research. This involved scaling up the model to interpret spectra of a hundred-fold larger size and using a wavelength encoding scheme adapted from the positional encodings in language models. Unlocking this important modality brings us closer to a foundation model that can be applied to any instrument in stellar astrophysics research.

\section{Transformer Foundation Model}

Unlike regular neural networks, Transformers are flexible in terms of input size and missing inputs~\citep{vaswani_attention_2017}. This flexibility is advantageous for developing a foundation model, as it allows the model to be adapted for a variety of tasks. LB23 developed a Transformer encoder-decoder model and trained it with 118 unique inputs,  consisting of various stellar properties and observations, with a maximum input size of \(N = 64\). They used a custom non-linear embedding process to encode tabular inputs:
\begin{equation}
    y_x = f(w_x\cdot M_x)+w_{b,x}
    \label{eq:embed}
\end{equation} with \(M_x\) as the value of the property `token', \(x\). \(w_x, w_{b,x}\) are learnable weights unique to \(x\), and \(f\) is a non-linear activation function. The architecture is structured such that you can provide the encoder whatever information you have about the star and then request any property by requesting vector \(w_x\) from the decoder. Our approach extends upon LB23 by introducing a wavelength encoding mechanism tailored for stellar spectroscopy with the capability to accept input spectra from across any range of wavelengths.

Spectroscopic observations consist of pixels of flux in specific wavelength bins, capturing the intensity of light within those wavelengths. There has been prior work using deep learning for APOGEE spectra to predict stellar properties and elemental abundances, notably with the \astronn\ framework \citep{leung_deep_2018}. It is a Convolutional Neural Network designed to work exclusively with APOGEE data and would not be useful when applied to other instruments like JWST or other wavelength ranges. In our experiments, we compare SpectraFM to \astronn. Unlike \astronn’s fixed input sizes and spectral regions, SpectraFM’s Transformer architecture allows for generalization and is designed for multi-instrument analysis.

In order to develop a model that can work with spectra from any instrument, we input every pixel in the spectra as an individual token rather than pre-processing the spectra with a fixed input size linear layer or PCA like other approaches that process spectra with Transformers \citep{zhang_spt_2024,zhang_maven_2024}. Each spectra token shares the embedding weights \(w_x\), \(w_{x,b}\), but are differentiated by a wavelength embedding added on to Equation~\ref{eq:embed} that consists of: 
\begin{equation}
\text{Wavelength Positional Encoding:} \quad \text{PE}(\hat{\lambda}, k) = 
\begin{cases} 
\sin\left(\frac{1000 \cdot \hat{\lambda}}{10000^{k/d_{\text{model}}}}\right), & \text{if } k \text{ is even} \\
\cos\left(\frac{1000 \cdot \hat{\lambda}}{10000^{k/d_{\text{model}}}}\right), & \text{if } k \text{ is odd} 
\end{cases}
\end{equation}
\begin{equation}
\hat{\lambda} = \frac{\lambda - \lambda_{\text{min}}}{\lambda_{\text{max}} - \lambda_{\text{min}}}, \quad k \in [0,d_\mathrm{model}-1]
\end{equation}
where \(d_\mathrm{model}=256\) embedding dimensions, and \(\lambda_\mathrm{min}=15,000\) \AA, \(\lambda_\mathrm{max}=17,000\) \AA  \,were chosen for infrared spectra ranges like the APOGEE dataset we use for this prototype, though can be expanded to a larger wavelength range in a future model.  This embedding scheme was inspired by \cite{rozanski_toward_2023} which used it in constructing a Transformer-based model for generating synthetic spectra. This allows for inputting spectra pixels of any wavelength into our model, such that it can trained on any instrument and wavelength range.

While LB23 focused on low-resolution Gaia spectra, SpectraFM's wavelength encoding mechanism pushes the model to recognize that spectra from different instruments should contain similar information about the star at the same wavelengths. In this study we demonstrate generalization from synthetic to real spectra, but this encoding is designed to support cross-instrument generalization as well.

\begin{figure}
  \centering
  \includegraphics[width=1.0\linewidth]{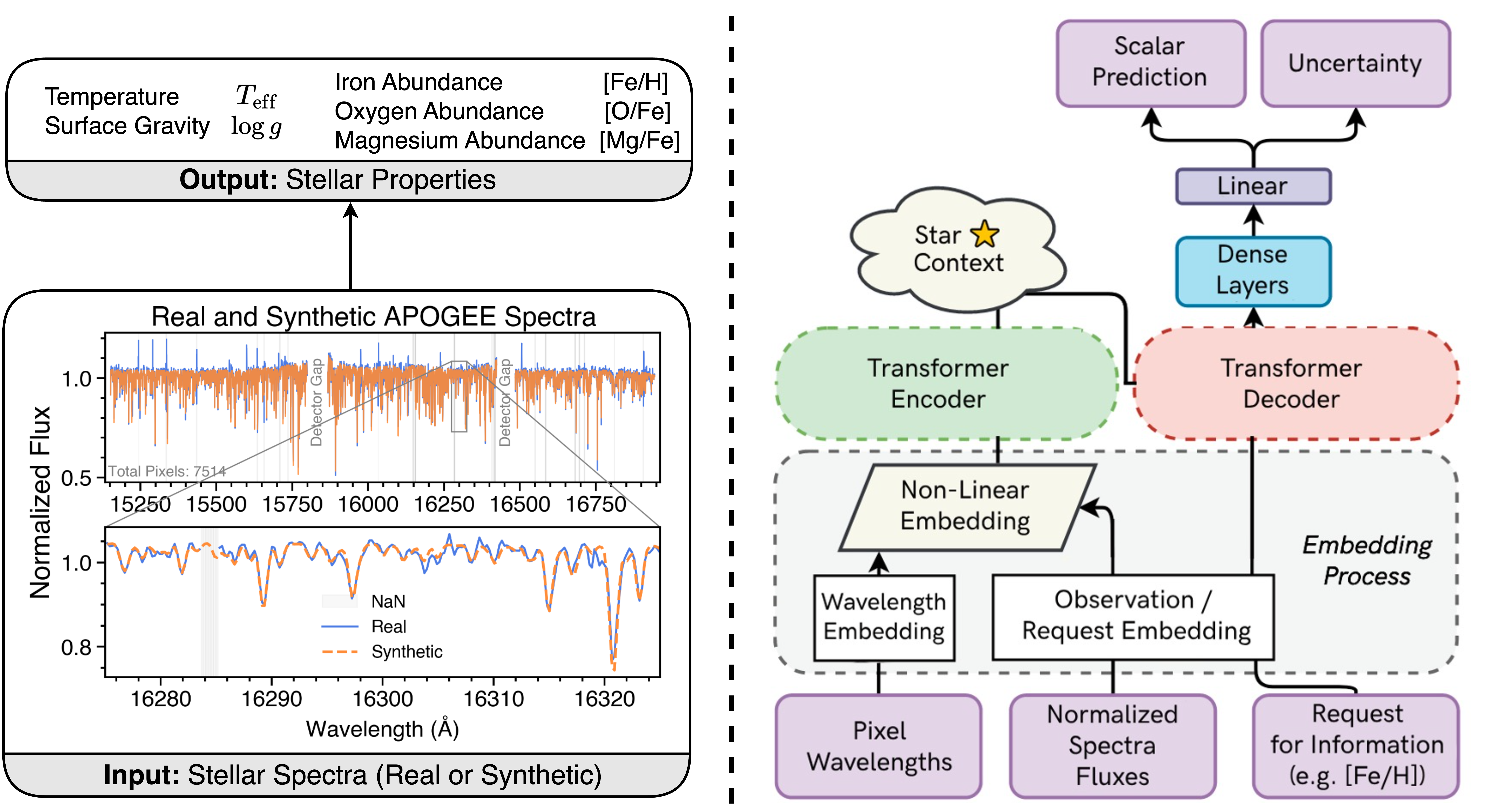}
  \caption{Overview of our Transformer-based foundation model architecture for stellar spectroscopy.
\textit{Left:} A comparison between the two possible inputs for a star: the real APOGEE infrared spectra and the synthetic best-fit spectra as generated by ASPCAP assuming simplified physics and without the observational issues that cause the noisy or NaN pixels in the real spectra. Also shown is the outputs requested from the model for each star that are important for astrophysical studies: the surface temperature and gravity, abundance of iron compared to hydrogen and abundance of oxygen and magnesium compared to iron. \textit{Right:} The encoder processes the spectral pixels as individual tokens, embedding both flux and wavelength. This allows for input spectra from any wavelength range and instrument. The encoded spectral data forms a context about the star, which the decoder uses, along with specific output requests, to generate predictions. The model also estimates a prediction confidence. Diagram adapted from LB23.}\label{fig:model}
\end{figure}

SpectraFM is an encoder-decoder Transformer model with approximately 8 million trainable parameters. Figure~\ref{fig:model} illustrates its architecture. The encoder contains two Transformer blocks interspersed with dense layers. The decoder receives the encoder output along with a request, encoded as the vector \(w_x\) representing the desired stellar property. The decoder consists of three  Transformer blocks and dense layers which result in the final prediction and a predicted uncertainty.

The loss function, adapted from LB23 and also used in \astronn, is a mean-squared loss that combines the uncertainties in both the training data and the model’s predictions. This approach allows the model to estimate a predictive uncertainty that reflects how confident it is in each prediction. The objective function \( L \) is defined as:

\[
L(y, \hat{y}) = \frac{(\hat{y} - y)^2}{2 e^{s}} + \frac{s}{2},
\]

where \( y \) and \( \hat{y} \) represent the ground truth and predicted values, respectively. The term \( s = \ln \left(\sigma_{\text{data}}^2 + \sigma_{\text{pred}}^2\right) \) includes both the known uncertainty in the data, \( \sigma_{\text{data}} \), and the predictive uncertainty of the model, \( \sigma_{\text{pred}} \), which is learned during training. Minimizing \( L \) not only improves model prediction accuracy but also improves the model's ability to produce a confidence measure for each prediction that accounts for variations that neither label uncertainties nor model predictions fully capture.

\section{Dataset and Training}

\begin{figure}
    \centering
    \includegraphics[width=1\linewidth]{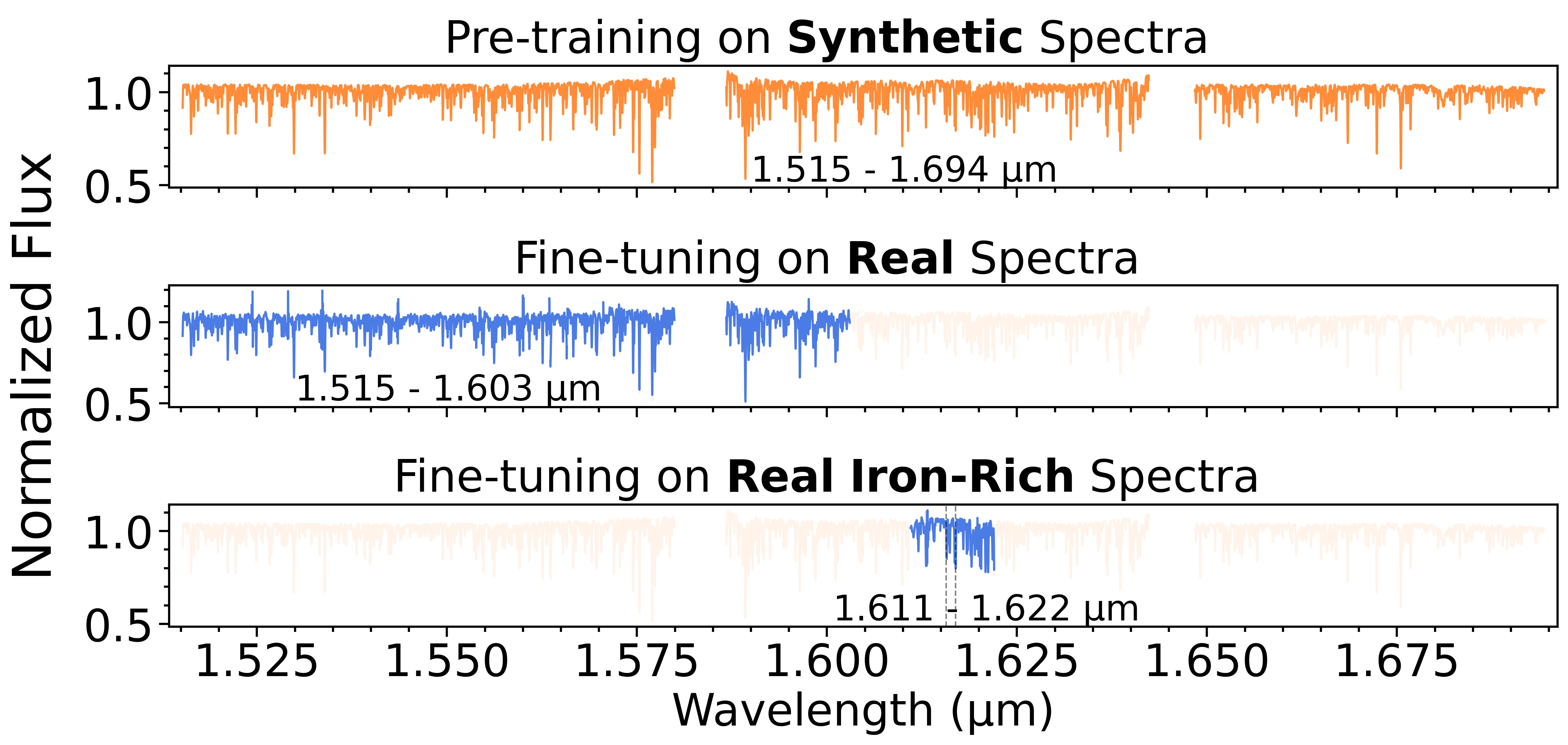}
    \caption{Wavelength regions used for each training and fine-tuning stage for this study. \textit{Top:} Synthetic spectra used for initial pre-training. \textit{Middle:} Real spectra used in the first fine-tuning step for comparison to \astronn. \textit{Bottom:} Real spectra used for the fine-tuning stage focusing on iron abundance prediction in a region not seen in the second step. This interval includes two prominent Fe lines and uses a limited training set of only 100 iron-rich stars to investigate transfer-learning.}
    \label{fig:training_regions}
\end{figure}

\begin{figure}
  \centering
  \includegraphics[width=\linewidth]{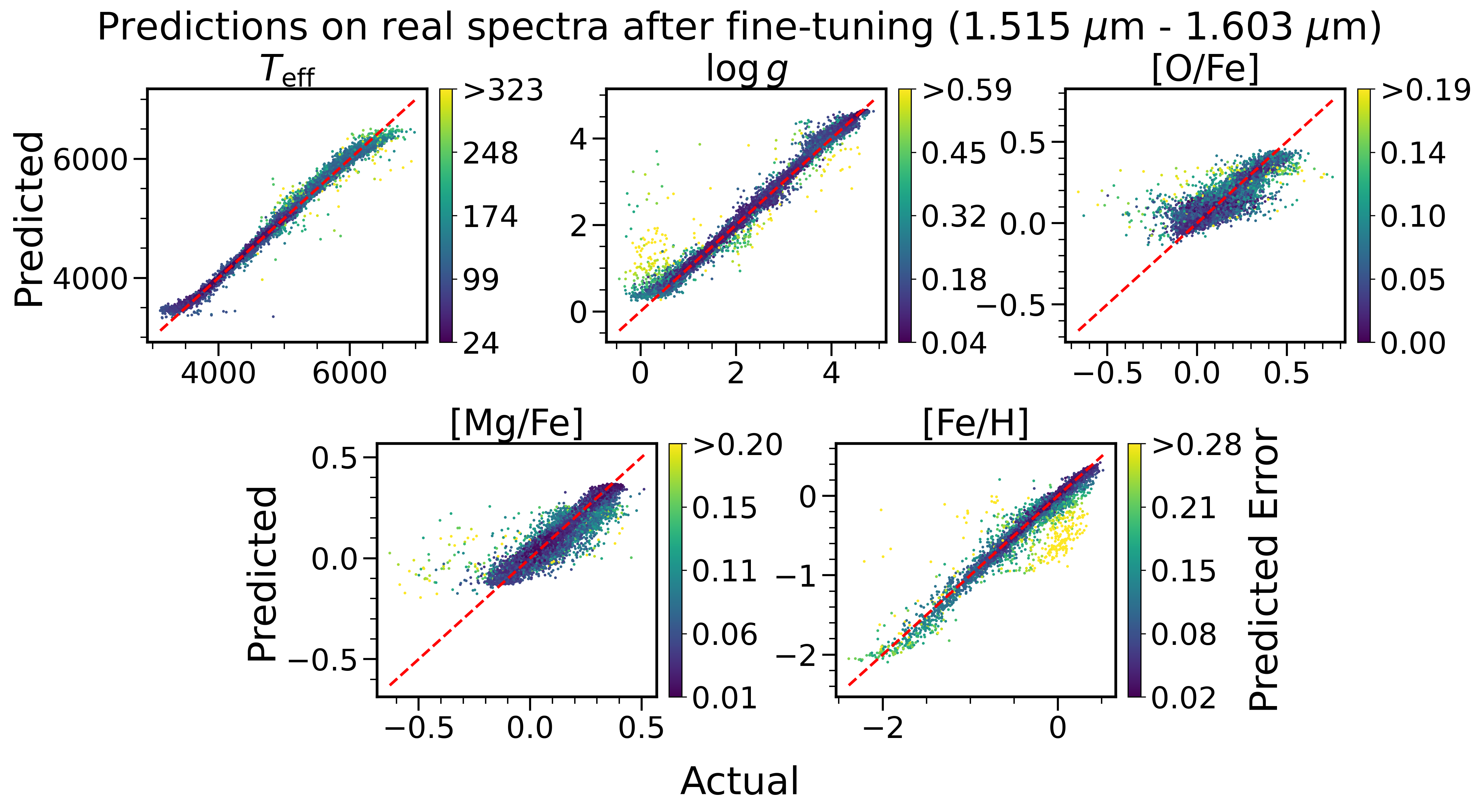}
  \caption{Stellar properties and chemical abundance predictions from real spectra using our model fine-tuned on real spectra. The point colors represent the prediction uncertainty learned by the base model. Our Transformer-based foundation model performs similarly to traditional deep learning methods like  \astronn~\citep{leung_deep_2018}}\label{fig:pt}
\end{figure}

In astrophysics, machine learning solutions often face challenges on tasks with limited availability of labeled data. For instance, while spectroscopic observations from a new telescope might be abundant, only a few dozen stars might have accurately measured properties like iron abundances ([Fe/H]). To address this, some approaches have trained machine learning models on synthetic spectra generated from theoretical simulations \citep{bialek_assessing_2020,fabbro_application_2018}. However, the synthetic gap often leads to inaccurate predictions. Our solution is to fine-tune on a small dataset of real spectra only after extensively pre-training on synthetic spectra.

The synthetic spectra used in this study come from simulations by ASPCAP (APOGEE Stellar Parameters and Abundances Pipeline) \citep{perez_aspcap_2016}. ASPCAP generates spectra under the assumption of 1D Local Thermodynamic Equilibrium (LTE), which simplifies the radiative transfer calculations in stellar atmospheres. For each real stellar spectra in APOGEE, ASPCAP releases the best-fit synthetic spectra. This synthetic spectra dataset thus has the same distribution of stellar properties and is split into a train/test set with its real spectra pairs to prevent test set leakage. The synthetic spectra are much cleaner than real observations, lacking instrumental noise and observational effects like cosmic rays, atmospheric effects and detector issues. Figure~\ref{fig:model} highlights the differences which lead to the `synthetic gap` that arises when training models only on synthetic spectra, which we attempt to resolve with fine-tuning on a small dataset of real spectra.

We select the following stellar properties for training: temperature (\teff), specific gravity (\logg), [Fe/H], [O/Fe], and [Mg/Fe]. \teff\ and \logg\ are important physical properties for classifying stars. [Fe/H], [O/Fe], and [Mg/Fe] abundances are essential for understanding the chemical and dynamical evolution of galaxies and were listed as top-priority elements for APOGEE to detect. Since Transformer models can work effectively with missing data, we do not need to remove stars from the training set with one or more missing properties. For more details about our data refinement pipeline see Section~\ref{sec:data_refinement_appendix}.

We pre-train our model on synthetic APOGEE spectra to predict combinations of \teff, \logg, [Fe/H], [O/Fe], and [Mg/Fe]. Our training set consists of \(\sim\)90k synthetic stars that matches the distribution of stars seen in APOGEE. Pre-training occurs on 4x Nvidia A100 GPUs over 325 epochs, taking approximately 21 hours. The learning rate starts at \(10^{-4}\) and varies throughout training according to a cyclic scheme known as Cosine Annealing with Warm Restarts, a common method that accelerates convergence \citep{loshchilov_sgdr_2017}. Specifically, we set the initial learning rate to \(10^{-4}\), the minimum learning rate to \(10^{-10}\), and the restart length to 50 epochs. We use the Adam optimizer \citep{kingma_adam_2017} for optimization. Our model is implemented using PyTorch \citep{paszke_pytorch_2019}. At this point, to compare to \astronn\ we fine-tune on a real spectra dataset of the same size but restrict it to the first half of the wavelength range to leave the second half for further tests focusing on transfer learning on an unseen wavelength range and fine-tuning on small datasets. The specific wavelength regions used at each step of training is illustrated in Figure~\ref{fig:training_regions}. The maximum input size of our model is limited to 512 pixels due to computational constraints. All predictions from inputs larger than 512 pixels are averages of predictions from 512 pixel chunks. For reference, the entire APOGEE spectra contain 7514 pixels.

Small training sets limit the accuracy of a machine learning algorithm, especially if the training set does not encompass the necessary label distribution. To mimic this scenario, we fine-tune again on a dataset limited to 100 iron-rich stars ([Fe/H] > -1) with a focus on [Fe/H] prediction. We choose the chunk of 512 pixels around two Fe lines in the second half of the spectra (1.611\(\mu\)m-1.622\(\mu\)m). For comparison, we train a fully connected neural network (FCNN) with 5 million trainable parameters, three hidden layers and dropout fraction of 10\% to reduce overfitting.

We suspect that any neural network trained solely on this limited dataset, e.g. convolutional neural networks like AstroNN or larger FCNNs, will fail in generalizing to iron-poor stars ([Fe/H] < -1), due to the out-of-distribution nature of the task. In this test, there is no exposure to iron-poor stars during training. SpectraFM benefits from pre-training on synthetic spectra, which enables it to transfer knowledge to real spectra and generalize beyond the fine-tuning training distribution. If our fine-tuned model accurately predicts [Fe/H] in the iron-poor range, this indicates that knowledge transferred from the synthetic pre-train and the fine-tuning on real spectra from a different wavelength range. We can compare the [Fe/H] prediction accuracy at each stage to see how knowledge is gained.

\begin{figure}
  \centering
  \includegraphics[width=\linewidth]{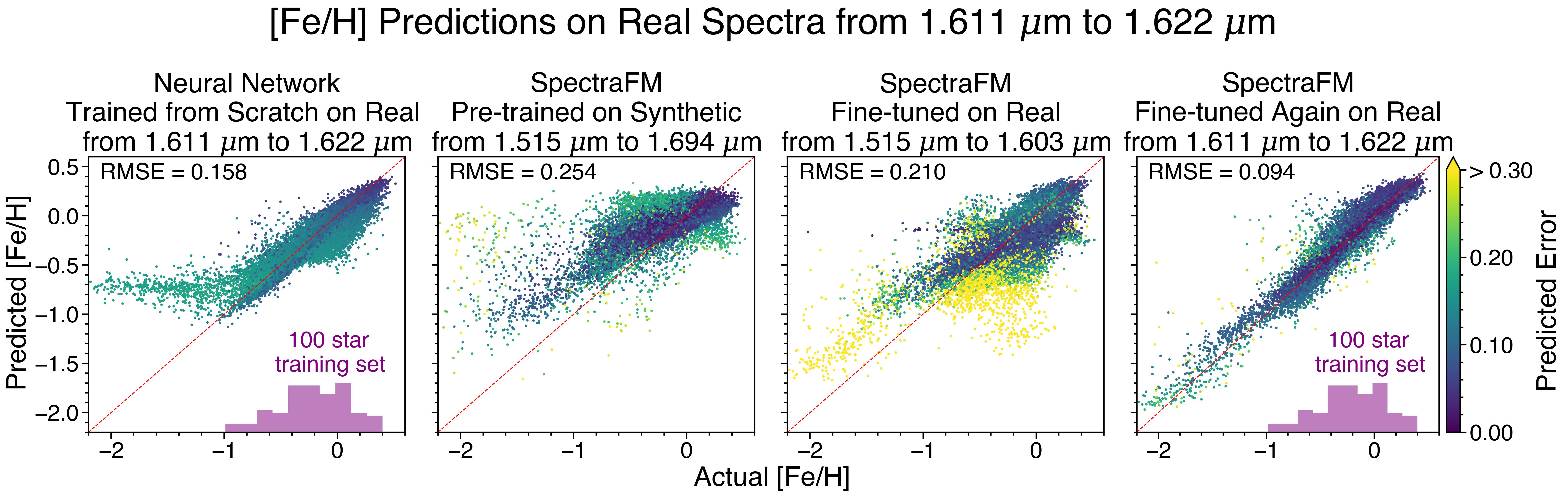}
  \caption{The [Fe/H] predictions from 512 pixels of real APOGEE spectra around two iron lines from the second half of APOGEE spectra (1.611 $\mu$m - 1.622 $\mu$m). \textit{Left}: a basic neural network trained on real APOGEE spectra in the target wavelength range with a dataset of only 100 iron-rich stars; \textit{Center-left}: the SpectraFM base model, pre-trained on synthetic spectra from approximately 90k stars;
  \textit{Center-right:} the SpectraFM base model fine-tuned on real APOGEE spectra from only the first half of the spectra which demonstrates increased performance even though it is not trained on real spectra in the target wavelength range;
  \textit{Right}: SpectraFM fine-tuned again on real APOGEE spectra in the target wavelength range with a dataset of only 100 iron-rich stars. Only SpectraFM, pre-trained on synthetic spectra and fine-tuned on 100 iron-rich real spectra, is able to generalize to iron-poor spectra for this task. This demonstrates that a pre-trained model is a better starting point than training from scratch for this small dataset task.}\label{fig:feh}
\end{figure}

\section{Results \& Discussion}

The prediction accuracy of our fine-tuned model on the first half of real spectra as seen in Figure~\ref{fig:pt} and Table \ref{tab:astroNN_vs_pretrainedFM} is similar to that of previous machine learning methods like \astronn\, \citep{leung_deep_2018}. Training only on the synthetic spectra is insufficient to handle real spectra due to the synthetic gap (middle column of Table \ref{tab:astroNN_vs_pretrainedFM}), while fine-tuning allows the model to make accurate predictions. We make a selection of $1.0 <$ \logg\,$< 3.5$ on our test set to match the \astronn\, training sample for a fair comparison.

\subsection{Transfer learning from synthetic to real data}

To investigate the extent to which the pre-training and the fine-tuning steps help the model generalize to previously unseen data regimes, we compare the results from our pre-trained and fine-tuned models to those from the basic neural network trained from scratch on the 100 iron-rich stars in Figure~\ref{fig:feh}. The from-scratch neural network fails to predict [Fe/H] in the iron-poor range and SpectraFM pre-trained on only synthetic spectra also performs poorly due to the simulations that generated the synthetic spectra not accurately modeling every physical process behind the real spectra (like in the middle column of Table \ref{tab:astroNN_vs_pretrainedFM}). We observe an increase in [Fe/H] accuracy from $\sigma_\mathrm{RMSE} = 0.254$ to $0.210$ by fine-tuning on real spectra in an entirely different wavelength range, which already outperforms the neural network trained from scratch in the iron-poor range ($\sigma_\mathrm{RMSE,[Fe/H]<-1.0} = 0.763$ vs. $0.510$). Fine-tuning again, but only using observed spectra of 100 iron-rich stars, which are the only real, observed spectra used for training in this wavelength region, leads to strong performance even in the iron-poor region ($\sigma_\mathrm{RMSE,[Fe/H]<-1.0} = 0.232$) and is the best overall estimator ($\sigma_\mathrm{RMSE}=0.094$). Skipping the first fine-tuning step and directly fine-tuning on the 100 iron-rich stars leads to similar performance: $\sigma_\mathrm{RMSE,[Fe/H]<-1.0} = 0.256$ and $\sigma_\mathrm{RMSE}=0.100$. Therefore, although fine-tuning on a different wavelength range with real spectra increased performance on this task compared to the base model, the knowledge required to achieve a high accuracy came from the 100 star fine-tune in the same wavelength range. Even though the model never sees iron-poor real spectra in the target wavelength range during training, it can make accurate predictions in this region, which demonstrates a new ability unlocked by generalizing knowledge from other tasks.

\begin{table}[t]
\footnotesize
\centering
\begin{tabular}{lccc}
\hline
Scatter & \astronn\, & SpectraFM Pre-trained Synthetic & SpectraFM Fine-tuned Real \\ \hline
\teff & 10 K & 183 K & 19 K \\
\logg & 0.037 dex & 0.326 dex & 0.056 dex \\
$[\mathrm{O/Fe}]$ & 0.020 dex & 0.038 dex & 0.021 dex \\
$[\mathrm{Mg/Fe}]$ & 0.015 dex & 0.034 dex & 0.019 dex \\
$[\mathrm{Fe/H}]$ & 0.011 dex & 0.048 dex & 0.019 dex \\ \hline
\end{tabular}
\caption{Comparison of prediction scatter for a selection of the test set between \astronn\, \citep{leung_deep_2018} and our foundation model after fine-tuning on real spectra. The scatter is the median absolute deviation: $\text{median}\left( \left| y_{\text{true},i} - y_{\text{pred},i} \right| \right)$. Our foundation model is only fine-tuned on the first half of the APOGEE spectra while \astronn\, was trained on the full spectra. Furthermore, \astronn\, likely included some of these stars in its training set. These stars satisfy $1.0 <$ \logg\,$< 3.5$ for a fair comparison since \astronn\ only trained on stars in this \logg\, range. The \astronn\ [O/Fe] and [Mg/Fe] are found from [X/Fe] = [X/H] - [Fe/H] since \astronn\ does not directly predict [X/Fe].}
\label{tab:astroNN_vs_pretrainedFM}
\end{table}

We found that freezing the parameters in the layers closer to the output of the last fine-tuned model, i.e. the Decoder and last layer of Encoder, led to stronger performance in the iron-poor region. This suggests that the layers closer to the output retain information about translating high-level spectral features to iron abundance from the synthetic pre-training. Meanwhile, the layers closer to the input require adjustment to recognize features in real spectra. This approach allows the model to leverage the pre-trained knowledge effectively without overfitting to the small real dataset. 

The ability to accurately predict [Fe/H] in the iron-poor range suggests that the synthetic-to-real knowledge transfer is successful and highlights the advantage of starting with a pre-trained foundation model, which requires only minimal adjustments to perform well across diverse conditions, such as new telescopes like JWST.

\subsection{Attention}
\label{sec:attention}

\begin{figure}
  \centering
  \includegraphics[width=\linewidth]{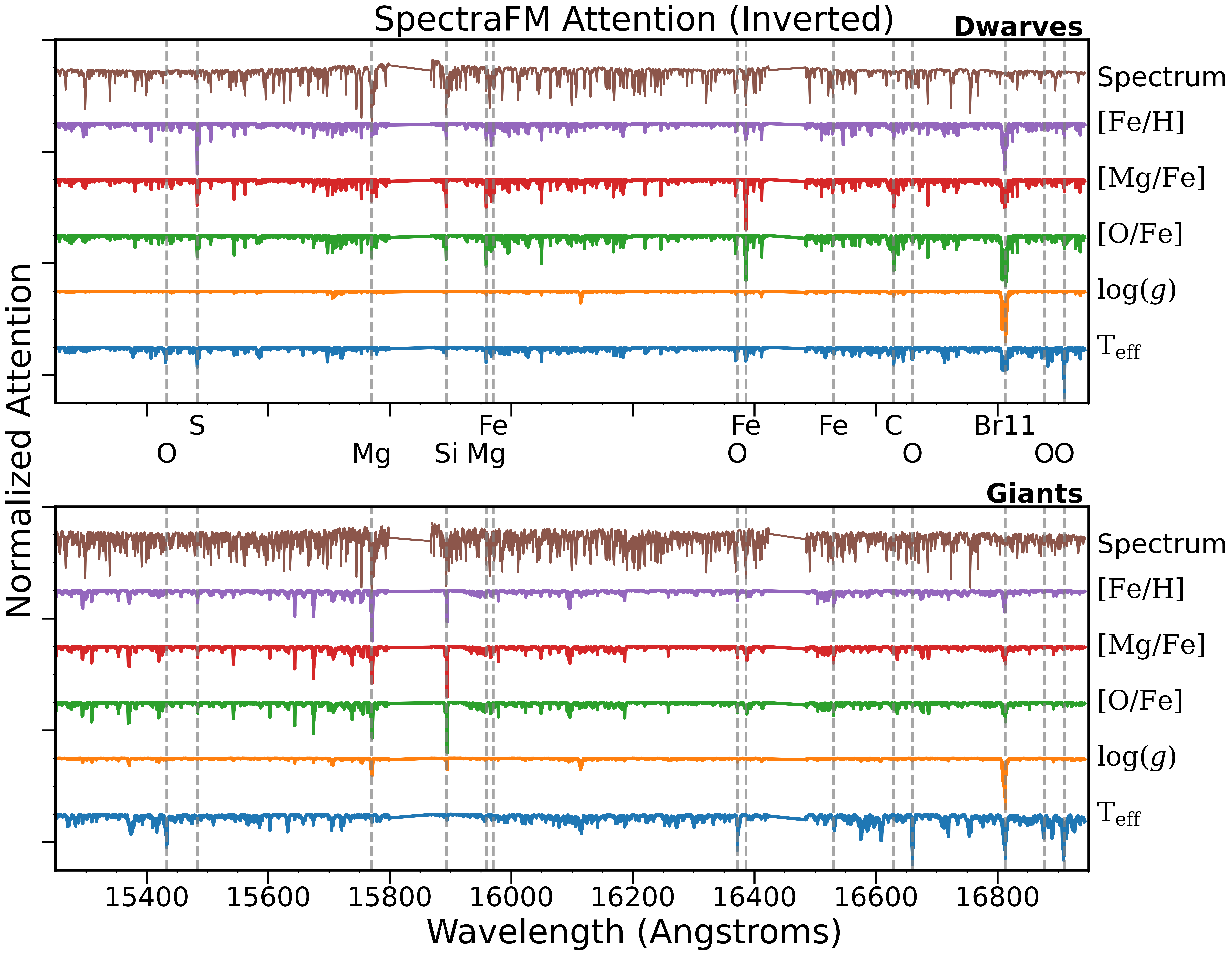}
  \caption{Inverted attention maps for synthetic spectra from dwarf (top) and giant (bottom) stars as analyzed by SpectraFM. Each row corresponds to a different stellar property: effective temperature (\teff), surface gravity (\logg), oxygen-to-iron ratio ([O/Fe]), magnesium-to-iron ratio ([Mg/Fe]), and iron abundance ([Fe/H]). The attention scores, normalized and averaged across stars, reveal the specific wavelength regions the model focuses on for each prediction. Spectral lines and wavelengths that contain information about certain elements are marked with dashed lines as determined by ASPCAP~\citep{perez_aspcap_2016}. The corresponding average spectrum for each stellar type is also plotted, to show how the attention aligns with physically meaningful spectral features. Br11 is a Hydrogen Brackett line~\citep{campbell_pre-main-sequence_2022} known to be sensitive to \logg. The attention maps have been vertically shifted for clarity. For further details on how attention scores are calculated and averaged, see Appendix~\ref{sec:appendix_attention}.}\label{fig:attention}
\end{figure}

A major concern in using machine learning for scientific research arises from the limited understanding of how a prediction is made and where the model sources the information to make that prediction. The attention mechanism behind Transformers can help us understand what information the model has learned to use \citep{vaswani_attention_2017}. We look at the relative values of attention scores that the Encoder transformer block of our base model assigns to each input pixel when making predictions for \teff, \logg, [Fe/H], [O/Fe], [Mg/Fe] (see Appendix~\ref{sec:appendix_attention} for more info on attention scores). We average the scores for giants and dwarfs, selected based on specific gravity (\logg\ < 3.0 for giants and \logg\ > 4.0 for dwarfs). The results of this can be seen in Figure~\ref{fig:attention}.

Stellar spectra contain many narrow absorption lines with physically meaningful information, so we invert the attention to easily compare the attention scores to the spectra and determine whether regions of high attention correspond to spectral features. The shape and depth of spectral lines contain information about the abundance of certain chemical elements along with the overall properties of the star, like \teff\ and \logg. The source of our simulated spectra, ASPCAP, identifies windows where the synthetic spectra are uniquely sensitive to the abundance of a given element \citep{jonsson_apogee_2020}. We find that many of the wavelengths the Transformer attends to hold physical significance. 

In particular, the attention for every property strongly focuses on the Br11 hydrogen line associated with a Brackett transition around 16813\AA\ \citep{campbell_pre-main-sequence_2022} known to be highly sensitive to the surface gravity \logg. Because determining the abundance of a chemical element from a stellar absorption line requires knowledge of the overall properties \teff\ and \logg, we expect the Br11 hydrogen line to be important for determining all elements and this is exactly what we see in Figure~\ref{fig:attention}. The [Fe/H] prediction looks at lines associated with iron at 16530\AA\ and 16386\AA, with the latter catching the attention of all other property predictions as well. \teff\ pays attention to O lines at 16660\AA\ and 16910\AA. The [Mg/Fe] and [O/Fe] attentions look similar, which is expected since Mg, O, Si, and Ca are all grouped as alpha elements, which form through the fusion of helium nuclei and are dispersed by core-collapse supernovae and so their abundances are usually correlated. Each of [Mg/Fe] and [O/Fe] pay strong attention to the Mg line at 15570\AA, Si line at 15893\AA, O line at 16372\AA\ and the Fe line at 16386\AA, which is necessary to determine their abundance relative to Fe.  

The attention mechanism effectively identifies and focuses on spectroscopically significant regions that correspond to chemical element transitions, demonstrating the model’s ability to learn physically meaningful features for accurate predictions. Previous machine learning methods like \astronn\ employed masking techniques to mitigate the risk of the model learning spurious correlations between different elements, which could introduce bias and compromise prediction accuracy. By selectively masking specific regions during training, \astronn\ focused on the relevant features for predicting a given element. A Transformer-based spectra foundation model has the advantage of a variable-length input and the ability to investigate attention. So future research could train the model to predict elements only based on their relevant features in the spectra, and then investigate the attention to ensure it is not basing its predictions off of correlations with other features that might not universally hold true. Furthermore, examining the model's attention when predicting a property may unveil hidden relationships not previously recognized, offering a new method for discovery.

\section{Conclusion}
This work presents a Transformer-based foundation model for stellar spectroscopy. The model, pre-trained on synthetic spectra and fine-tuned with limited real data, outperforms traditional neural networks by bringing out-of-distribution tasks inside the distribution with pre-training. Its attention mechanism targets key physical features in the spectra, ensuring predictions are physically grounded.

Our results show that for new tasks in astrophysics, fine-tuning a foundation model will likely lead to better results than a basic neural network. For example, if a James Webb Space Telescope data release contained a limited number of stars with measured abundances, our results suggests that fine-tuning our foundation model on this training set would lead to a highly accurate model that could then be used to get abundances for other stars. Our model could be fine-tuned to predict other properties as well, for example stellar ages, mass, and spectra-photometric distances \citep{leung_simultaneous_2019, leung_variational_2023}. Our understanding of the Galaxy's evolutionary history, like the formation of the bar, disk, and stellar halo, along with our understanding of globular clusters and dwarf galaxies, rely heavily on measuring these properties to high accuracy \citep[e.g.,][]{leung_measurement_2023}.

Future directions include integrating diverse datasets to enhance cross-instrument and cross-domain generalization. We plan to exploit our model's flexibility by pre-training on all major stellar spectroscopic surveys such as LAMOST DR9 (10 million stars, 370-900 nm)~\citep{liu_lamost_2020}, GALAH DR3 (588k stars, optical and infrared bands) \citep{buder_galah_2021}, and Gaia DR3 low-resolution spectra (220 million stars, 330-1050 nm)~\citep{de_angeli_gaia_2022}, each with differing resolutions. A wider variety of training data should increase performance on all tasks due to knowledge generalization.

Few-shot learning is also an area of interest, especially for its applications for analyzing rare stellar types and datasets from new instruments that are too small for fine-tuning. Moving to a decoder-only approach with positional encoding would enable such ability.

This work lays the groundwork for developing comprehensive astronomical foundation models, which could greatly assist in data analysis in large-scale surveys.

For reproducibility and transparency, we open-source our code, training scripts, and models for SpectraFM at \url{https://github.com/NolanKoblischke/SpectraFM_NeurIPS_FM4Science}.

\begin{ack}
We thank Henry W. Leung for useful discussion. NK and JB acknowledge financial support from the Natural Sciences and Engineering Research Council of Canada (NSERC; funding reference number RGPIN-2020-04712). NK is partially funded through the NSERC Canada Graduate Scholarship - Master’s.

This research was enabled in part by support provided by Compute Ontario (\url{https://www.computeontario. ca}) and the Digital Research Alliance of Canada (\url{alliancecan.ca}).

Funding for the Sloan Digital Sky Survey IV has been provided by the Alfred P. Sloan Foundation, the U.S. Department of Energy Office of Science, and the Participating Institutions. SDSS-IV acknowledges support and resources from the Center for High Performance Computing at the University of Utah. The SDSS website is \url{https://www.sdss.org/}.

This work has made use of data from the European Space Agency (ESA) mission \textit{Gaia} (\url{https://www.cosmos.esa.int/gaia}), processed by the Gaia Data Processing and Analysis Consortium (DPAC, \url{https://www.cosmos.esa.int/web/gaia/dpac/consortium}). Funding for the DPAC has been provided by national institutions, in particular the institutions participating in the \textit{Gaia} Multilateral Agreement.
\end{ack}

{
\small
\bibliography{references}
}


\appendix

\section{Attention Mechanism in Transformers}
\label{sec:appendix_attention}

The attention mechanism enables the model to focus on the relevant spectral features for each prediction. In the model, the self-attention is computed as:

\[
\text{Attention}(Q, K, V) = \text{softmax}\left( \frac{QK^T}{\sqrt{d_k}} \right) V
\]

where \(Q\), \(K\), and \(V\) represent the query, key, and value matrices derived from the input tokens (spectral pixels)~\citep{vaswani_attention_2017}. The attention scores \( \alpha_{ij} \) between query \(Q_i\) and key \(K_j\) are defined as:

\[
\alpha_{ij} = \frac{\exp\left(\frac{Q_i \cdot K_j}{\sqrt{d_k}}\right)}{\sum_{k} \exp\left(\frac{Q_i \cdot K_k}{\sqrt{d_k}}\right)}
\]

These normalized scores determine the relative importance of each key-value pair. We present attention scores from the second layer of the encoder, as these higher layers capture more complex patterns rather than simple structures~\citep{raghu_vision_2022}. This is analogous to convolutional neural networks, where deeper layers detect meaningful features (e.g. the shape of spectral lines) while lower layers identify simpler structures (e.g., edges or peaks). Our attention analysis in Section~\ref{sec:attention} aims to determine if the model identifies physically relevant wavelengths and the information stored in spectral lines at these wavelengths. The attention scores are averaged across all attention heads and across all stars within each category (dwarfs or giants) to smooth out individual variations and emphasize the key regions that the model consistently focuses on.

\section{Data Refinement for APOGEE DR17 Spectra}
\label{sec:data_refinement_appendix}
To prepare a high-quality training dataset, we applied a series of selections to the APOGEE DR17 spectroscopic dataset~\citep{majewski_apache_2017}. First, we selected stars with a signal-to-noise ratio above 200, ensuring reliable measurements. Next, we removed stars flagged for quality issues and observational problems. We then eliminated binaries by filtering stars with high radial velocity scatter (\(v_\mathrm{scatter}<1\)km/s) since this often indicates the source is actually a binary star. For stars with multiple observations, we deduplicated the data by selecting the highest-SNR observation for each star. These steps reduced the dataset from 733,901 spectra to 128,762. Our synthetic spectra dataset is sourced from the best-fit synthetic spectra for each of these stars found by ASPCAP~\citep{perez_aspcap_2016}. We divided the refined set into 70\% for training, 20\% for testing, and 10\% for training validation. Stars with NaN labels remained in the dataset, as our loss function and Transformer model handles missing data effectively during training.

\end{document}